\documentclass{ws-ijmpcs}
\usepackage{url,overcite}
\usepackage{xfrac}

\newcommand{\Vek}[1]{\boldsymbol{#1}}
\newcommand{\fract}[2]{{\textstyle\frac{#1}{#2}}} 

\renewcommand{\draftnote}{}
\renewcommand{\catchline}{}
\renewcommand{\openaccesstext}{}

\begin{document}

\markboth{H. Weigel, J.P. Blanckenberg}
{Heavy baryons with strangeness ...}

\catchline{}{}{}{}{}
%
%%%%%%%%%%%%%%%%%%%%%%%%%%%%%%%%%%%%%%%%%%%%%%%%%%%%%%%%%%%%%%%%%%%%

\title{Heavy baryons with strangeness in soliton models}

\author{H. Weigel, J. P. Blanckenberg}

\address{Physics Department, Stellenbosch University \\ 
Matieland 7602, South Africa}

%\begin{history}
%\received{Day Month Year}
%\revised{Day Month Year}
%\published{Day Month Year}
%\end{history}

\maketitle

\begin{abstract}
We present some recent results from soliton model calculations 
for the spectrum of baryons with a single heavy quark. The model
comprises chiral symmetry for light flavors and (approximate) heavy 
spin--flavor symmetry for the heavy quarks. We focus on flavor symmetry 
breaking for strangeness degrees of freedom.
\end{abstract}

\ccode{PACS numbers: 11.30.Rd, 12.39.Dc, 12.39.Hg, 14.20.-c}

\section{Introduction}	
Baryons containing heavy quarks have drawn renewed attention since the 
potential discovery of pentaquark states with hidden charm\cite{Aaij:2015tga}.
Here we present a model calculation for the spectrum of baryons with a 
single heavy quark. 

We start from a chiral soliton model with pseudoscalar and vector meson fields that 
reasonably well describes the spectrum, static properties and meson nucleon scattering. 
These studies are reviewed in Ref.\refcite{Weigel:2008zz}. We then augment this model 
by coupling mesons with a single heavy quark (charm or bottom). This coupling is 
constructed to reproduce the heavy spin--flavor symmetry as the mass of the heavy
meson is sent to infinity\cite{Schechter:1992ue}. The soliton then produces an
attractive potential for the heavy mesons and the resulting bound states are central 
to the investigation of heavy baryons\cite{Schechter:1995vr}. In this presentation, 
which is mainly based on Ref.\refcite{Blanckenberg:2015dsa}, we particularly discuss
how flavor symmetry breaking between the light non--strange and strange quarks is 
included when coupling the chiral soliton and the heavy meson bound state to form a 
baryon with a heavy quark.

\section{The Model}
We start from a chiral soliton model for light baryons. The major building block is 
the non--linear representation of the pesudoscalar mesons in form of the chiral 
field~$U$. Rather than using higher derivative terms as in the Skyrme 
model~\cite{Skyrme:1961vq,Adkins:1983ya} to stabilize the soliton, here stabilization 
is achieved by coupling the pseudoscalar mesons to the vector mesons $\rho$ and 
$\omega$~\cite{Jain:1987sz,Schwesinger:1988af,Park:1991fb}.
Though we utilize a three flavor model, the soliton profiles are embedded in 
the $SU(2)$ subgroup of isospin:
\begin{equation}
U_0(\Vek{r})={\rm exp}\left[\Vek{\tau}\cdot\hat{\Vek{r}}F(r)\right]\,,\quad
\omega^{(0)}_\mu(\Vek{r})=\omega(r)\,g_{\mu0}\,,\quad
\Vek{\rho}^{(0)}(\Vek{r})=\Vek{\tau}\times\hat{\Vek{r}}\,\frac{G(r)}{r}\,.
\label{eq:soliton}
\end{equation}
The profile functions $F(r)$, $\omega(r)$ and $G(r)$ are determined by minimizing 
the classical energy, $E_{\rm cl}$ subject to boundary conditions that ensure unit 
winding number for the mapping of coordinate space into flavor space. 
Note that, on this classical level, the spatial components of the isoscalar 
field $\omega_\mu$ and the time components of isovector field~$\Vek{\rho}_\mu$ 
are zero. We generate states with 
good baryon numbers by introducing\footnote{Also  profile functions are induced 
for the spatial components of $\omega_\mu$ and the time components of 
$\rho_\mu$~\cite{Meissner:1986js,Park:1991fb}.} and quantizing collective
coordinates for the flavor orientation $A(t)\in SU(3)$
\begin{equation}
U(\Vek{r},t)=A(t) U_0(\Vek{r}) A^\dagger(t)
\quad {\rm and} \quad
\Vek{\tau}\cdot\Vek{\rho}_\mu(\Vek{r},t)=
A(t)\Vek{\tau}\cdot\Vek{\rho}^{(0)}_\mu(\Vek{r})A^\dagger(t)\,.
\label{eq:cc1}
\end{equation}
Defining eight angular velocities $\Omega_a$ 
\begin{equation}
\frac{i}{2}\sum_{a=1}^8\Omega_a\lambda_a=
A^\dagger(t)\,\frac{dA(t)}{dt}\,,
\label{eq:cc2}
\end{equation}
enables a compact form for the collective coordinate Lagrange function 
that arises from the light meson fields 
\begin{equation}
L_l(\Omega_a)=-E_{\rm cl}+\frac{1}{2}\alpha^2\sum_{i=1}^3\Omega_i^2
+\frac{1}{2}\beta^2\sum_{\alpha=4}^7\Omega_\alpha^2
-\frac{\sqrt{3}}{2}\Omega_8\,.
\label{eq:cc3}
\end{equation}
The last term is only linear in the time derivative and originates from the
Wess--Zumino--Witten action~\cite{Wi83}. The coefficients $\alpha^2$ and $\beta^2$ 
are radial integrals of the profile functions and represent
moments of inertia for rotations in isospace and the strangeness subspace of 
flavor $SU(3)$, respectively. The form of the collective coordinate Lagrangian, 
Eq.~(\ref{eq:cc3}) is generic to all chiral soliton models. The particular 
numerical values for the classical energy and the moments of inertia are, of course, 
subject to the particular model for which we will follow Ref.\refcite{Park:1991fb}.
The rotations also induce vector field components that vanish classically (see above).
So far the collective coordinates only enter through their time derivatives; $A$ does 
not appear explicitly as we have not yet included $SU(3)$ flavor symmetry breaking 
contributions.

\section{Heavy Meson Bound States}

In the heavy flavor limit the pseudoscalar ($P$) and vector meson ($Q_\mu$) components
become degenerate\cite{Neubert:1993mb} and must therefore be combined in a single 
multiplet $H=\fract{1}{2}\left(i\gamma_5P+\gamma^\mu Q_\mu\right)$. The Dirac spinor 
labels are subject to the heavy flavor transformation properties while the light flavor 
index of this multiplet ($P$ and $Q_\mu$ are $SU(3)$ flavor spinors) is coupled to the 
light meson fields according to rules of chiral symmetry. This coupling to the soliton 
induces a potential for $P$ and $Q_\mu$ that allows for bound state solutions with energy 
eigenvalue $\omega$. Since the soliton configuration itself has non--zero orbital angular 
momentum the most strongly bound solution has a P--wave structure in the pseudoscalar 
component:
\begin{align}
P&=\frac{{\rm e}^{i\omega t}}{\sqrt{4\pi}}\Phi(r) 
\hat{\Vek{r}}\cdot \hat{\Vek{\tau}}\chi\, ,\qquad
Q_0=\frac{{\rm e}^{i\omega t}}{\sqrt{4\pi}}\Psi_0(r)\chi \cr
Q_i&=\frac{{\rm e}^{i\omega t}}{\sqrt{4\pi}}\left[i\Psi_1(r){\hat r}_i
+\fract{1}{2}\Psi_2(r)\epsilon_{ijk}{\hat r}_j\tau_k\right]\chi\,.
\label{eq:pansatz}
\end{align}
where $\chi$ is a constant three component spinor. Since the coupling to the light
mesons occurs via a soliton in the isospin subspace, only the first two components
of $\chi$ are non--zero. The parameterization that emerges by left multiplication with 
$\hat{\Vek{r}}\cdot\hat{\Vek{\tau}}$ and, of course, has different profile functions,
leads to the S--wave bound states. 

Since the heavy meson fields are spinors in $SU(3)$ flavor space, they are also 
subject to collective flavor rotation from Eq.~(\ref{eq:cc1}),
\begin{equation}
P\,\longrightarrow\, A(t) P \qquad {\rm and}\qquad
Q_\mu\,\longrightarrow\, A(t) Q_\mu\,,
\label{eq:hcc1}
\end{equation}
where the right hand sides contain the bound state profile functions. From this we
get the collective coordinate Lagrange function from the heavy fields
\begin{equation}
L_h(\Omega_a)=-\omega\chi^\dagger \chi 
+\frac{1}{2\sqrt{3}}\chi^\dagger\Omega_8 \chi
+\rho\chi^\dagger\left(\Vek{\Omega}\cdot\frac{\Vek{\tau}}{2}\right)\chi\,.
\label{eq:hcc2}
\end{equation}
The coefficient $\rho$ in the last term is a radial integral over the profile 
functions\cite{Harada:1997we}. Again, the flavor rotation matrix $A$ does not 
appear explicitly.

\section{Symmetry Breaking and Mass formula}

So far we have not taken into account three flavor symmetry breaking as reflected 
by the different (current) quark masses of up, down and strange quarks, $m_u$, $m_d$ 
and $m_s$, respectively. While it is appropriate to work with $m_u=m_d$, 
the deviation $m_s\gg m_u,m_d$ is substantial and must be included. 
It is incorporated in the effective chiral Lagrangian by adding terms like
\begin{equation}
\mathcal{L}_{\rm sb}\sim \frac{f_\pi^2m_\pi^2}{4}{\rm Tr}\left[
\begin{pmatrix}1 & 0 & 0 \cr 0 & 1 & 0 \cr 0 & 0 & x \end{pmatrix}
\left(U+U^\dagger-2\right)\right]+\ldots
\qquad {\rm where} \qquad
x\,\widehat{=}\,\frac{2m_s}{m_u+m_d}\gg1\,.
\label{eq:lsb1}
\end{equation}
Eventually they describe different masses and decay constants of strange and 
non--strange mesons\footnote{Symmetry breaking for the heavy mesons, proportional 
to {\it e.g.} $M_{B_s}^2-M_B^2$, is also included. For brevity the corresponding 
Lagrangian is not displayed here.}. Once the profile functions are substituted 
and the spatial integration is performed, symmetry breaking induces an explicit 
$A$ dependence in the collective coordinate Lagrange function
\begin{equation}
L_{\rm sb}=-\frac{x}{2}\widetilde{\gamma}\left[1-D_{88}(A)\right]
\quad {\rm with} \quad
D_{ab}=\fract{1}{2}{\rm Tr}\left[\lambda_aA\lambda_bA^\dagger\right]\,.
\label{eq:lsb2}
\end{equation}
Again, $\widetilde{\gamma}$ is a radial integral\footnote{The notation is 
chosen to distinguish it from $\gamma=x\widetilde{\gamma}$ in the
literature\cite{Weigel:2008zz}.} over all profile 
functions\cite{Blanckenberg:2015dsa}. 

Collecting pieces from Eqs.~(\ref{eq:cc1},\ref{eq:hcc2}) and~(\ref{eq:lsb2}) 
and Legendre transforming for the angular velocities via the
right $SU(3)$ generators $R_a=\frac{\partial L}{\partial \Omega_a}$
yields the Hamilton operator whose eigenvalues are the baryon masses. 
This results in the mass formula
\begin{align}
E&=E_{\rm cl}+\left(\frac{1}{\alpha^2}-\frac{1}{\beta^2}\right)\frac{r(r+1)}{2}
+\frac{\epsilon(x)}{2\beta^2}
-\frac{3}{8\beta^2}\left(1-\frac{N}{3}\right)^2\cr\cr
&\hspace{2cm}
+|\omega|N+\frac{\rho}{2\alpha^2}\left[j(j+1)-r(r+1)\right]N\,,
\label{eq:massf}
\end{align}
where $N$ counts the number of heavy quarks. Here $\epsilon(x)$ is 
the eigenvalue of 
$O_{\rm sb}=\sum_{a=1}^8R_a^2+x\beta\widetilde{\gamma}\left[1-D_{88}(A)\right]$
according to the Yabu--Ando approach\cite{Yabu:1987hm} subject to the 
constraint $R_8=\sqrt{3}(1-\fract{N}{3})/2$. For heavy baryons this 
constraint then requires diquark $SU(3)$ representations. Furthermore $j$ is 
the total spin of the consider baryon and $r(r+1)$ is the eigenvalue of 
$\sum_{i=1}^3R_i^2$. It is zero and one for the anti--symmetric and the symmetric 
diquark wave--functions, respectively\footnote{Non--heavy baryons have 
$N=0$ and $r=j$}. We stress that obtaining the eigenvalues $\epsilon(x)$ 
of the operator $O_{\rm sb}$ amounts to a non--perturbative treatment of 
symmetry breaking in the light flavor sector.

\section{Results}

All coefficients in the mass formula, Eq.~(\ref{eq:massf}) are determined
from the soliton model calculation detailed in 
Refs.\refcite{Park:1991fb,Schechter:1995vr} and \refcite{Harada:1997we}. 
The only free parameter is the strength, $x$ of flavor symmetry breaking. 
It has been estimated\cite{Gasser:1982ap,Schechter:1992iz,Harada:1995sj} 
from meson properties to be in the range $x\sim20..30$. In tables \ref{tab:cp}, 
\ref{tab:cn} and \ref{tab:bp} we list our predictions for the mass 
differences\footnote{We concentrate on mass differences to avoid ambiguities 
from the vacuum polarization energy\cite{Meier:1996ng}.} of the heavy baryons 
with the respect to the nucleon and the $\Lambda_{c,b}$ and compare them to 
empirical data\cite{Agashe:2014kda}. The mass differences within a given heavy 
quark sector is overestimated. For example 
$M_{\Omega_c}-M_{\Lambda_c}=463{\rm MeV}$ for $x=25$, while the empirical value 
is $409{\rm MeV}$. Further increase of $x$ worsens the picture. On the other 
hand, a sizable value ($x\sim30$) for the symmetry breaking is required for 
a good agreement for non--heavy baryons\cite{Park:1991fb}. The splitting 
between different heavy sectors is predicted on the low side: when compared
to the nucleon, the $\Lambda_c$ and $\Lambda_b$ are about $100{\rm MeV}$ 
and $300{\rm MeV}$ too low, respectively. This is inherited from the heavy 
flavor calculation which overestimates\cite{Schechter:1995vr} the 
binding energies in the sense that it is too close to the estimate from exact 
heavy flavor symmetry.  This can also be seen from the parity splitting which 
is underestimated by about $50{\rm MeV}$ (it vanishes in the heavy limit). 
Together with the effect of $SU(3)$ symmetry breaking the overestimated binding 
combines to acceptable agreement for the mass differences between the double 
strange baryons $\Omega_c$ and $\Omega_b$ and the nucleon, at least for $x=30$.
It has been argued~\cite{Harada:1997we} that kinematical corrections 
({\it e.g.} substituting the reduced mass in the bound state problem) due 
to the soliton not being infinitely heavy change the predicted bound state
energies appropriately. 

For $j=\sfrac{1}{2}$ and positive parity the observed mass difference between 
$\Sigma$ and $\Xi$ decreases and even changes sign when the heaviest flavor 
turns from strange via charm to bottom: $M_{\Xi}-M_{\Sigma}=125,14,-17{\rm MeV}$. 
This is (partially) reflected by our calculation, {\it e.g.} for $x=25$ we find 
the mass differences $101$, $23$ and $6{\rm MeV}$. Since the hyperfine splitting 
only has a moderate effect, we find a similar scenario for the negative 
parity channel and it will be interesting to compare it with future data.

\begin{table}[t]
\tbl{Predicted mass differences for the positive parity
baryons with a single charm quark for two different strengths of
$SU(3)$ symmetry breaking in comparison with experimental
data (PDG)\cite{Agashe:2014kda}. The mass differences
with respect to the nucleon and the $\Lambda_c$ are denoted by
$\Delta_N$ and $\Delta_{\Lambda_c}$, respectively.}
{\begin{tabular}{c|c||c|c|c|c||c|c}
&&\multicolumn{2}{|c}{$x=25$}&
\multicolumn{2}{|c||}{$x=30$}&
\multicolumn{2}{|c}{Data (PDG)}\cr
\hline
Bary. & $(I,j,r)$ & $\Delta_N$ & $\Delta_{\Lambda_c}$
& $\Delta_N$ & $\Delta_{\Lambda_c}$
& $\Delta_N$ & $\Delta_{\Lambda_c}$
\cr \hline
$\Lambda_c$ & $(0,\sfrac{1}{2},0)$
& 1230 & 0 & 1233 & 0 & 1347 & 0  \cr
$\Sigma_c$ & $(1,\sfrac{1}{2},1)$
& 1423 & 193 & 1425 & 192 & 1515 & 168 \cr
$\Xi_c$ & $(\sfrac{1}{2},\sfrac{1}{2},0)$
& 1446 & 216 & 1486 & 253 & 1529 & 182 \cr
$\Omega_c$ & $(0,\sfrac{1}{2},1)$
& 1693 & 463 & 1756 & 523 & 1756 & 409 \cr
$\Xi_c$ & $(\sfrac{1}{2},\sfrac{1}{2},1)$
& 1557 & 328 & 1588 & 355 & 1637 & 290 \cr
$\Sigma_c$ & $(1,\sfrac{3}{2},1)$
& 1464 & 234 & 1466 & 233 & 1579 & 232 \cr
$\Xi_c$ & $(\sfrac{1}{2},\sfrac{3}{2},1)$
& 1598 & 369 & 1629 & 396 & 1706 & 359  \cr
$\Omega_c$ & $(0,\sfrac{3}{2},1)$
& 1734 & 504 & 1797 & 564 & 1831  & 484
\end{tabular}
\label{tab:cp}}
\end{table}

\begin{table}[b]
\tbl{Same as table \ref{tab:cp} for the
negative parity charmed baryons.}
{\begin{tabular}{c|c||c|c|c|c||c|c}
&&\multicolumn{2}{|c}{$x=25$}&
\multicolumn{2}{|c||}{$x=30$}&
\multicolumn{2}{|c}{Data (PDG)}\cr
\hline
Bary. & $(I,j,r)$ & $\Delta_N$ & $\Delta_{\Lambda_c}$
& $\Delta_N$ & $\Delta_{\Lambda_c}$
& $\Delta_N$ & $\Delta_{\Lambda_c}$
\cr \hline
$\Lambda_c$ & $(0,\sfrac{1}{2},0)$
&1479 & 249 & 1482 & 249 & 1653 & 306  \cr
$\Sigma_c$ & $(1,\sfrac{1}{2},1)$
&1664 & 434 & 1666 & 433 & - & - \cr
$\Xi_c$ & $(\sfrac{1}{2},\sfrac{1}{2},0)$
&1695 & 465 & 1735 & 502 & 1851 & 504 \cr
$\Omega_c$ & $(0,\sfrac{1}{2},1)$
&1934 & 704 & 1997 & 764 & - & - \cr
$\Xi_c$ & $(\sfrac{1}{2},\sfrac{1}{2},1)$
&1798 & 569 & 1829 & 596 & - & - \cr
$\Sigma_c$ & $(1,\sfrac{3}{2},1)$
&1717 & 487 & 1719 & 486 & - & - \cr
$\Xi_c$ & $(\sfrac{1}{2},\sfrac{3}{2},1)$
&1851 & 622 & 1882 & 649  & - & - \cr
$\Omega_c$ & $(0,\sfrac{3}{2},1)$
&1987  & 757 & 2050 & 817  & - & -
\end{tabular} \label{tab:cn}}
\end{table}

\begin{table}[h]
\tbl{Same as table \ref{tab:cp} for the
positive parity bottom baryons. Here~$\Delta_{\Lambda_b}$ is
the mass difference with respect to $\Lambda_b$.}
{\begin{tabular}{c|c||c|c|c|c||c|c}
&&\multicolumn{2}{|c}{$x=25$}&
\multicolumn{2}{|c||}{$x=30$}&
\multicolumn{2}{|c}{Data (PDG)}\cr
\hline
Bary. & $(I,j,r)$ & $\Delta_N$ & $\Delta_{\Lambda_b}$
& $\Delta_N$ & $\Delta_{\Lambda_b}$
& $\Delta_N$ & $\Delta_{\Lambda_b}$
\cr \hline
$\Lambda_b$ & $(0,\sfrac{1}{2},0)$
& 4391 & 0 & 4394 & 0 & 4681 & 0 \cr
$\Sigma_b$ & $(1,\sfrac{1}{2},1)$
& 4601 & 210 & 4603 & 209 & 4872 & 191 \cr
$\Xi_b$ & $(\sfrac{1}{2},\sfrac{1}{2},0)$
& 4608 & 216 & 4647 & 253 & 4855 & 174 \cr
$\Omega_b$ & $(0,\sfrac{1}{2},1)$
& 4871 & 480 & 4935 & 540 & 5110 & 429 \cr
$\Xi_b$ & $(\sfrac{1}{2},\sfrac{1}{2},1)$
& 4736 & 345 & 4766 & 372 & - & - \cr
$\Sigma_b$ & $(1,\sfrac{3}{2},1)$
& 4617 & 226 & 4619 & 225 & 4893 & 212 \cr
$\Xi_b$ & $(\sfrac{1}{2},\sfrac{3}{2},1)$
& 4751 & 360 & 4782 & 387 & 5006 & 325 \cr
$\Omega_b$ & $(0,\sfrac{3}{2},1)$
& 4887 & 496 & 4950 & 556 & - & -
\end{tabular}\label{tab:bp}}
\end{table}

\section{Summary}

We have presented model predictions for the spectrum of baryons with a single
heavy quark. Our analysis culminates in a single mass formula for the spectrum.
The ingredients of the mass formula are the binding energies of the heavy meson,
the flavor symmetry breaking among the non--heavy flavors and the hyperfine
splitting. In these aspects our study builds on and extends an earlier chiral 
soliton model approach\cite{Momen:1993ax} on heavy baryons with strangeness.
We stress that, except for a moderate uncertainty of the light symmetry breaking
strength, all parameters in our mass formula are obtained from a single 
model Lagrangian. Though not presented here, we recall that this model also 
predicts\cite{Park:1991fb} the spectrum and static properties of light baryons 
with reasonable agreement to empirical data.

\section*{Acknowledgments}

One of us (HW) is grateful to the organizers for providing this worthwhile
conference. This work supported in parts by the NRF under grant~77454.

\bibliographystyle{ws-ijmpcs}

\begin{thebibliography}{10}

\bibitem{Aaij:2015tga}
R.~Aaij {\em et~al.}, {\em Phys. Rev. Lett.} {\bf 115}, 072001  (2015).

\bibitem{Weigel:2008zz}
H.~Weigel, {\em Lect.Notes Phys.} {\bf 743}, 1  (2008).

\bibitem{Schechter:1992ue}
J.~Schechter and A.~Subbaraman, {\em Phys.Rev.} {\bf D48}, 332  (1993).

\bibitem{Schechter:1995vr}
J.~Schechter, A.~Subbaraman, S.~Vaidya and H.~Weigel, {\em Nucl.Phys.} {\bf
  A590}, 655  (1995).

\bibitem{Blanckenberg:2015dsa}
{J. P. Blanckenberg and H. Weigel},
{\em Phys. Lett.} {\bf B750}, 230 (2015).

\bibitem{Skyrme:1961vq}
{T.\ H.\ R.\ Skyrme}, {\em Proc.Roy.Soc.Lond.} {\bf A260}, 127  (1961).

\bibitem{Adkins:1983ya}
{G.\ S.\ Adkins, C.\ R.\ Nappi and E.\ Witten}, {\em Nucl.Phys.} {\bf B228},
  552  (1983).

\bibitem{Jain:1987sz}
{P.\ Jain, R. Johnson, Ulf-G.\ Mei{\ss}ner, N.\ W. Park and J.\ Schechter},
  {\em Phys.Rev.} {\bf D37}, 3252  (1988).

\bibitem{Schwesinger:1988af}
{B. Schwesinger, H. Weigel, G. Holzwarth and A. Hayashi}, {\em Phys. Rept.}
  {\bf 173}, 173  (1989).

\bibitem{Park:1991fb}
{N.\ W.\ Park and H.\ Weigel}, {\em Nucl.Phys.} {\bf A541}, 453  (1992).

\bibitem{Meissner:1986js}
{Ulf-G. Mei{\ss}ner, N. Kaiser and W. Weise}, {\em Nucl.Phys.} {\bf A466},
  685  (1987).

\bibitem{Wi83}
E.~Witten, {\em Nucl. Phys.} {\bf B223}, {422, 433}  (1983).

\bibitem{Neubert:1993mb}
M.~Neubert, {\em Phys.Rept.} {\bf 245}, 259  (1994).

\bibitem{Harada:1997we}
M.~Harada, A.~Qamar, F.~Sannino, J.~Schechter and H.~Weigel, {\em Nucl.Phys.}
  {\bf A625}, 789  (1997).

\bibitem{Yabu:1987hm}
H.~Yabu and K.~Ando, {\em Nucl.Phys.} {\bf B301}, 601  (1988).

\bibitem{Gasser:1982ap}
J.~Gasser and H.~Leutwyler, {\em Phys.Rept.} {\bf 87}, 77  (1982).

\bibitem{Schechter:1992iz}
J.~Schechter, A.~Subbaraman and H.~Weigel, {\em Phys.Rev.} {\bf D48}, 339
  (1993).

\bibitem{Harada:1995sj}
M.~Harada and J.~Schechter, {\em Phys.Rev.} {\bf D54}, 3394  (1996).

\bibitem{Meier:1996ng}
F.~Meier and H.~Walliser, {\em Phys. Rept.} {\bf 289}, 383  (1997).

\bibitem{Agashe:2014kda}
{K.\ A.\ Olive, {\it et al.}}, {\em Chin.Phys.} {\bf C38}, 090001  (2014).

\bibitem{Momen:1993ax}
A.~Momen, J.~Schechter and A.~Subbaraman, {\em Phys.Rev.} {\bf D49}, 5970
  (1994).

\end{thebibliography}

\end{document}